\documentclass[twocolumn,showpacs,aps,prl,superscriptaddress]{revtex4}

\usepackage{graphicx}
\usepackage{amsmath}
\usepackage{latexsym}

\begin{document}

\title{Possible observation of phase coexistence \\ 
of the $\nu=1/3$ fractional quantum Hall liquid and a solid}

\author{G.A. Cs\'athy}
\affiliation{Department of
Electrical Engineering, Princeton University, Princeton, NJ 08544}

\author{D.C. Tsui}
\affiliation{Department of
Electrical Engineering, Princeton University, Princeton, NJ 08544}

\author{L.N. Pfeiffer}
\affiliation{Bell Labs, Lucent Technologies, Murray Hill, NJ 07974}

\author{K.W. West}
\affiliation{Bell Labs, Lucent Technologies, Murray Hill, NJ 07974}

\date{\today}

\begin{abstract}

We have measured the magnetoresistance of a very low density 
and an extremely high quality two-dimensional hole system. 
With increasing magnetic field applied perpendicularly
to the sample we observe the sequence of
insulating, $\nu=1/3$ fractional quantum Hall liquid, 
and insulating phases. 
In both of the insulating phases in the vicinity of the $\nu=1/3$
filling the magnetoresistance has 
an unexpected oscillatory behavior with the magnetic field.
These oscillations are not of the Shubnikov-de Haas type
and cannot be explained by spin effects. 
They are most likely the consequence of the formation of a
new electronic phase which is intermediate between
the correlated Hall liquid and a disorder pinned solid.

\end{abstract}
\pacs{73.43.-f, 73.20.Qt, 73.40.-c, 73.63.Hs}
\maketitle

Two-dimensionally confined charge carriers subjected to
low temperatures ($T$) and high magnetic fields ($B$)
display a multitude of phases. 
Among them are the series of well-known fractional quantum Hall (FQH) liquids
\cite{fraq} that terminate at high $B$ in the
high field insulating phase \cite{insul_e,insul_h}.
There is mounting evidence \cite{peide} that
the high field insulating phase is crystalline, often thought to
be the Wigner solid (WS). The terminal FQH state,
the liquid state with the lowest Landau level filling factor $\nu$,
is found to be the $\nu=1/5$ in 2D electron \cite{insul_e}
and the $\nu=1/3$ in 2D hole systems \cite{insul_h}.
Exception is the electron gas confined to an extremely narrow
quantum well, a system for which the terminal FQH state is at
$\nu=1/3$ \cite{conf}.
The difference between electron and hole samples is attributed to a 
profound change of the ground state energies of the FQH liquid and the WS
that is due to Landau level mixing (LLM) \cite{insul_h}. LLM occurs
when the separation between the single particle Landau levels
is significantly less than the Coulomb interaction energy
\cite{yoshioka}. Since the effective mass $m^*$ of the holes is
about 5 times that of the electrons, holes have a smaller
cyclotron energy $\hbar e B / m^*$ and therefore larger LLM.
This enhanced LLM is thought to change the terminal FQH state.

Besides forming at high $B$ field due to magnetic quenching 
of the kinetic energy, the WS can also form
at $B=0$ when the ratio $r_s$ of the
Coulomb and the Fermi energies is large
\cite{wigner}. The interaction parameter $r_s$ can be expressed as
$r_s = m^* e^2/(4 \pi \epsilon \hbar^2 \sqrt{\pi p})$, with $p$ the
areal density of the charges.
It was thought that the WS, the ground state at large $r_s$, melts 
into the Fermi liquid around $r_s=37$ \cite{tanatar}. 
However, exact diagonalization on small systems \cite{prichard} 
with the Coulomb interaction included
revealed that the melting of the WS into a Fermi liquid
with decreasing $r_s$ could occur in two
steps resulting in the intriguing possibility of an intermediate phase
between the WS and the Fermi liquid. Similarly, the possibility
of an intermediate phase in charges confined to 2D
has been shown for a system
in which the Coulomb interaction is screened by a
nearby  metallic gate \cite{spivak1,spivak2} 
and for another system that exhibits a first order transition in 
the presence of non-uniformly distributed dopants when at least one of the 
phases is insulating \cite{ky}.
The intermediate state could be liquid-crystal-like 
\cite{spivak2,liqcryst}  or an admixture of interpenetrating 
liquid and solid phases \cite{prichard,spivak1,spivak2,ky}. 

In this Letter we report on a study of an extremely dilute 
two-dimensional hole (2DH)
system with the density of $p=0.98 \times 10^{10}$ cm$^{-2}$. 
At the 38~mK base temperature of our refrigerator 
the $\nu=1/3$ FQH state is intercalated between
two insulating phases. To our surprise, in both of these low and 
high field insulating phases we observe oscillations of the 
magnetoresistance with the $B$ field, with a period of several mT. 
The observed oscillations may be the consequence of an intermediate phase
of the 2DH system formed by coexisting FQH liquid and crystal phases.
This intermediate phase is similar to the intermediate phase
predicted in the $B=0$ case, with the exception that the
crystal coexists with the $\nu=1/3$ FQH liquid, rather than
the $B=0$ Fermi liquid.

Our samples are grown on a (311){\it A} GaAs substrate. 
The 2DH system is formed 
in a 30~nm wide GaAs/AlGaAs quantum well with Silicon dopants 
on both sides of the well. The
mobility of the holes is $\mu=0.36 \times 10^6$ cm$^2$/Vs at 38~mK.
A recent cyclotron resonance experiment on a different piece
from the same wafer reported $m^*=0.37$ in units of free electron
mass \cite{keji} which together with
the very low density of $p=0.98 \times 10^{10}$ cm$^{-2}$
yields an exceptionally high $r_s=30$.
The ohmic contacts to the hole gas are made of InZn alloy.
The size of the samples and the distance between the ohmic contacts
are of the order of 1~mm and the voltage probes are
on the side cut along the [$\bar{2}$33] direction. 
Transport measurements were carried out 
using the low frequency lock-in technique at an excitation current of 1~nA.

\begin{figure}[t]
\begin{center}
\includegraphics[width=3.9in]{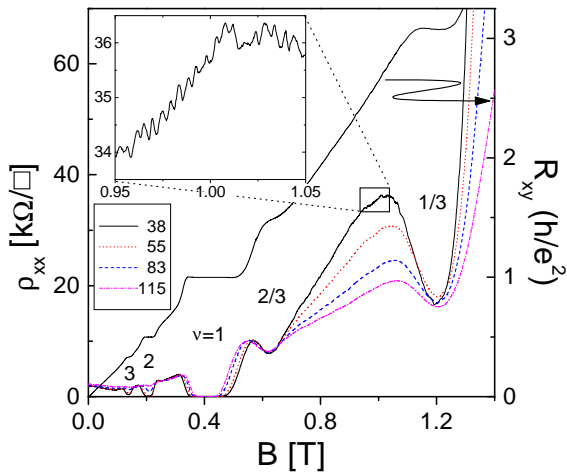}
\end{center}
\caption{\label{f1}
The longitudinal resistivity $\rho_{xx}$
and the Hall resistance R$_{xy}$ at 38~mK. Legend shows the temperatures
for $\rho_{xx}$ in mK. 
}
\end{figure}

In Fig.1 we show the dependence on $B$ of $\rho_{xx}$ 
at four different temperatures and the Hall resistance at 38~mK. 
We observe fully developed 
integer quantum Hall states at $\nu=1$, 2, and 3 
and partially developed $\nu=2/3$ 
and $1/3$ FQH states. As we cool the sample,
$\rho_{xx}$ at $\nu=1/3$ initially increases and only
at the lowest temperatures $T<60$~mK starts decreasing.
In contrast, $\rho_{xx}$ at $\nu=2/3$ is decreasing
with decreasing $T$ in our measurement range.
At $\nu<1/3$ or equivalently for $B>1.22$~T 
we observe the insulating phase that has been associated with 
the weakly pinned WS \cite{insul_h,peide}. 
While in 2DH samples of $p=4 \times 10^{10}$ cm$^{-2}$ at 
$\nu=0.37$ a reentrant insulating phase 
with $\rho_{xx} > 300$~k$\Omega/\Box$ has been reported
\cite{insul_h}, in our sample we 
observe no clear reentrant behavior.
The resistance between $1/3<\nu<2/3$ 
is a weakly insulating one and it reaches a maximum of only 
36~k$\Omega/\Box$ at $\nu=0.40$ at 38~mK. 

From the lowest $T$ curve of Fig.1 it is apparent that $\rho_{xx}$ 
has a fine structure in the vicinity of its maximum close to 
$B=1.0$~T. The inset of Fig.1 shows a magnified view of this region.
Surprisingly, there are oscillations in $\rho_{xx}$. 
The amplitude and period of these oscillations do not depend on 
the direction of the sweep of the $B$ field, the sweep rate of the field,
and the current as long there are no heating effects (for $<$2~nA).
The oscillations are not seen for all voltage probing contacts
and we have observed
them in three different samples from the same wafer.
They are present at different cool downs, though their amplitude 
and the phase exhibit small variations. 

\begin{figure}[b]
\begin{center}
\includegraphics[width=4.0in]{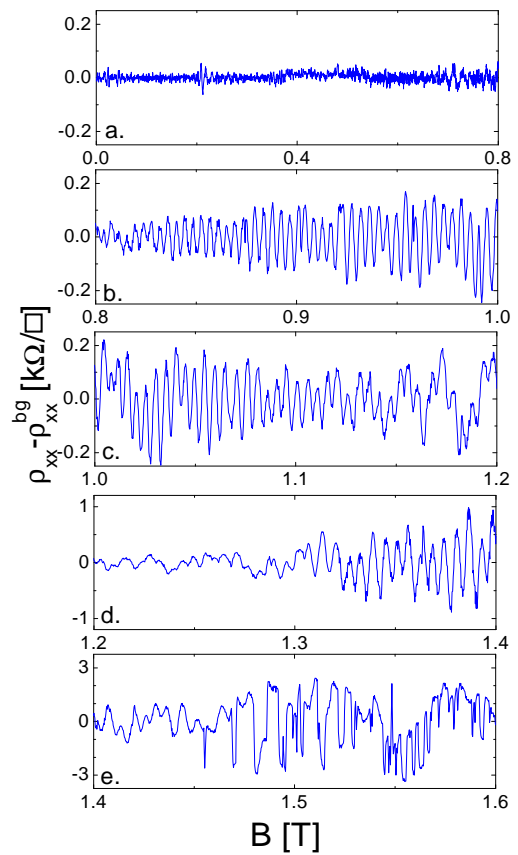}
\end{center}
\caption{\label{f2}
The dependence of $\rho_{xx}$ on $B$ after the subtraction of a smooth
background $\rho_{xx}^{bg}$. Note the change of scale of the
vertical axis for panels d. and e.
}
\end{figure}

To show the oscillations over the whole $B$ range,
we subtracted from $\rho_{xx}$ a background $\rho_{xx}^{bg}$
that is a slowly varying function of $B$. This background  is
obtained by fitting $\rho_{xx}$ over about 30 periods
of oscillations to polynomials of degree 7. The result is shown in Fig.2. 
While the oscillations cannot be observed for $B<0.79$~T,
they are present in both the low and the high field insulating
phases. Their amplitude reaches a maximum at $B=1.0$~T
in the low field insulating phase, it almost
vanishes at $\nu=1/3$, then it is very large again beyond 1.3~T
in the high field insulating phase. The oscillations persist
to fields as high as 1.45~T, a value beyond which there are
large nonperiodic fluctuations.
In a first order approximation the amplitude of 
the oscillations scales with $\rho_{xx}$. 

\begin{figure}[t]
\begin{center}
\includegraphics[width=3.2in]{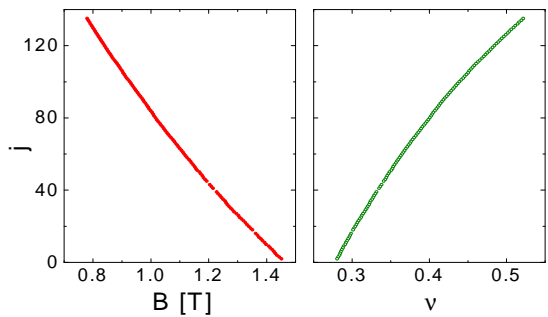}
\end{center}
\caption{\label{f3}
The dependence on $B$ and $\nu$
of the index $j$ of the maxima of the oscillations of $\rho_{xx}$.
}
\end{figure}

In the following we investigate the nature of the observed oscillations.
Fig.3 shows the index $j$ labeling the local maxima of $\rho_{xx}$ 
as a function of $B$ and $\nu=hp/(eB)$.
$j$ versus $\nu$ is not linear, therefore 
the oscillations are not of the Shubnikov-de Haas type.
Replacing the abscissa with $1/(B-B_{\nu=1/2})$ 
leads to even stronger nonlinearities (not shown), therefore 
the oscillations cannot be due to composite Fermions \cite{jain}.
Since the $\nu=1/3$ state is known to be spin-polarized,
it is unlikely that the oscillations are due to a spin effect.
$j$ as a function of $B$, also shown in Fig.3, is quasi-linear. 
The oscillations of $\rho_{xx}$ are therefore quasi-periodic in $B$
with the period being the inverse slope of the $j$ versus $B$ curve.
The resulting period $\delta B$, shown in Fig.4a,
increases linearly with $B$. This period is found to be $T$-independent.

Periodic modulation of the magnetoresistance of the FQH effect
can result from two mechanisms. One of them is the 
Aharonov-Bohm (AB) interference of the electronic wave
function between two conductive paths \cite{ab}. 
Due to constructive and destructive interference
the resistance is periodic with successive
penetration of one quantum of the magnetic flux $\Phi_0=h/e$ through
the area enclosed by the two paths. 
A typical period of 6~mT measured in our samples
corresponds to (0.83~$\mu$m)$^2$.
Our samples, however, are of the oder of 1~mm and 
do not have any {\it intentional} patterning
on scale of 1~$\mu$m. Inhomogeneities of the order of 1~$\mu$m can arise
either from the depletion of the 2DH due to defects in the
host GaAs/AlGaAs crystal or from an inhomogeneous electronic
phase consisting of two interpenetrating phases.
A second mechanism for oscillatory magnetoresistance
is the periodic transfer of 
elementary charges between two interpenetrating phases. 
In the following we will discuss these possibilities.

The AB effect around fixed defects
has been previously observed in lithographically etched
rings \cite{Chang,goldman,kurdak}, in a 
micron size Hall bar \cite{simmons}, in
singly connected geometries \cite{Kouven,harris}, and
in particular, it has been demonstrated in hole samples \cite{pepper,yau}.
Depleted regions in our samples due to defects in the host crystal 
could similarly lead to AB interference.
We argue that this scenario is unlikely. 
First, for the lithographically etched samples in Refs.
\cite{Chang,harris,goldman,pepper}
the AB oscillations are present at filling at least up to $\nu=2$.
Similarly, quasiperiodic resistance fluctuations
due to resonant tunneling through states that are magnetically
bound to defects have been observed for $\nu \leq 4$ filling
in a 2~$\mu$m wide Hall bar \cite{simmons}.
We do not observe any of the oscillations for $\nu>0.52$
and, in particular, near any integer fillings.
Second, the period of AB interference in artificially
created mesoscopic systems \cite{Chang,goldman,harris,pepper}
is independent of $B$. Indeed, in a submicron size square
\cite{Chang}, in a quantum antidot electrometer \cite{goldman}, and
in a quantum point contact \cite{harris} the period of
the AB oscillations changes less than 5\% over a $B$ range of
more than 100 periods. In contrast, the period $\delta B$ in our sample
almost doubles in the 0.8-1.4~T range. Third, it is improbable
that growth defects generated in three different samples
would roughly be of the same size yielding
oscillations of similar period and amplitude.
These results therefore suggest that the AB effect
around a fixed defect in the host crystal cannot explain our data.

\begin{figure}[b]
\begin{center}
\includegraphics[width=4.2in]{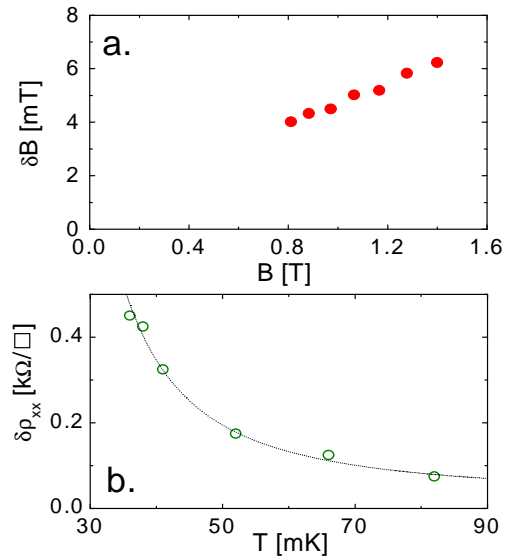}
\end{center}
\caption{\label{f4}
The period $\delta B$ of the oscillations of the longitudinal
resistivity as a function of $B$ (panel a.)
and the amplitude of the oscillations
$\delta \rho_{xx}$ at $B=1$~T as a function of $T$ (panel b.).
The dotted line is a guide to the eye.}
\end{figure}

A second possibility for AB interference is when
the 2DH system is not homogeneous but it
phase separates on the scale of 1~$\mu$m into two coexisting phases.
Since the oscillations are present in the insulating phases
on both sides of $\nu=1/3$, the $\nu=1/3$ FQH liquid
and a crystalline phase are natural choices for the two phases.
In this scenario, in the vicinity of the $\nu=1/3$ filling there is
an intermediate phase which consists of droplets
of FQH liquid and patches of crystal.
The patches of crystal are localized by the disorder present
in the sample and the droplets of correlated FQH liquid,
that percolate through the sample, support the AB interference.
The intermediate phase is similar to the theoretically
predicted $B=0$ intermediate phase described in the introduction
\cite{prichard,ky,spivak1,spivak2} with the FQH liquid
replacing the $B=0$ Fermi liquid.
Since increasing $r_s$ results in enhanced LLM that
in turn decreases the energy difference between
the solid and the FQH liquid \cite{yoshioka}, it is possible that
the extremely large $r_s=30$ of our sample causes
vanishingly small energy difference between the two phases.
Under such conditions both phases are equally favored
and the ground state is an intermediate phase as a consequence of
the competition between the two phases.
From an experimental viewpoint, such a situation
is plausible because, due to the almost $T$-independent $\rho_{xx}$
at $\nu=1/3$, the FQH liquid is extremely fragile.

Details of our data, in this interpretation, are determined by
the delicate balance between the coexisting phases.
The oscillations  of $\rho_{xx}$ are quite sharp with a well
defined period. This is most likely because the AB patches
have a very narrow size distribution, perhaps due to surface tension effects.
However, we cannot discard the possibility that only one
patch is dominant in determining the period.
The linearly increasing $\delta B$ with $B$ we measure
is consistent with the liquid phase being enclosed by the interfering paths.
Indeed, a droplet of $N_{liq}$ holes and of area
$A_{liq}$ must shrink with increasing $B$ since the filling
factor the liquid $\nu=N_{liq} \Phi_0 /(A_{liq} B)$ remains constant. 
A decreasing area with increasing $B$ results in an increasing period 
via the AB condition $A_{liq}\delta B = \Phi_0$.
Furthermore, combining the last two equations 
we obtain $\delta B=\nu B/N_{liq}$. The linearly increasing period with $B$
agrees with the data of Fig.4a and yields $N_{liq} = 66$.
The $T$-independence of $\delta B$ can be explained
by the rapid destruction with increasing $T$ of the phase coherence
while droplets have not significantly changed with $T$.
Lastly we note, that AB interference over distances of 1~$\mu$m is possible
in our sample because the AB effect has been observed in samples
with linear size close to the elastic scattering length \cite{kurdak}
and this length in our sample is 0.59~$\mu$m.

The alternative route that leads to periodic
modulation of $\rho_{xx}$ is the charge transfer between
the two electronic phases earlier described. 
We argued that the size of a liquid droplet 
shrinks with increasing $B$. At a certain value of $B$ 
it could become more favorable for the droplet
to exchange an elementary charge with a patch of solid
rather than shrink. As a result, the
magnetic flux of the droplet changes by $3 \Phi_0$
and the droplet relaxes size after the charge transfer. 
With further increase of $B$, the size of the droplet 
shrinks again and the described process is periodic with $B$. 
It is interesting to point out that for this process, as
opposed to the AB effect, it is
possible that the period $\delta B$ of the charge transfer
is independent of the size of the droplets 
and consequently the oscillations are not smeared out by averaging.
We wish to point out that the 
charge transfer process between the two phases described here 
and the charging of isolated islands, known as the Coulomb blockade 
\cite{cb}, are quite different. While both processes involve
charge transfer, the former one is between the coexisting compressible
solid and incompressible correlated liquid phases 
in 2D, the latter one is in between
two charge reservoirs through the 0D states of the
quantum dot between the reservoirs.

In conclusion, we have observed quasiperiodic modulation
of the Shubnikov-de Haas 
magnetoresistance in very low density and large $r_s$ 2DH samples. 
These oscillations are present in the insulating regions on 
both sides of the $\nu=1/3$ FQH state. We argue, that
due to the large LLM the 2DH gas most likely is not homogeneous, instead 
it phase separates into droplets of FQH liquid and patches of solid.
Assuming that the oscillations observed are the manifestation of the 
AB effect, the size of the patches around which the interference
occurs is of the order of 1~$\mu$m. 
Charge transfer between the liquid and solid phases might also
provide a viable explanation.

We thank Boris Spivak and Kun Yang for useful discussions. This research
was funded by the DOE and the NSF.

\end{document}